\documentclass[aps,prd,reprint,superscriptaddress,bibnotes,amsmath,amssymb,longbibliography]{revtex4-2}
\usepackage{graphicx}
\usepackage{booktabs}
\usepackage[hidelinks]{hyperref}

\begin{document}
\title{A Paired Monte Carlo Intensity-Ratio Method for Incorporating Detector Resolution in Amplitude Analysis}
\author{Zhenxuan Li}
\email[Electronic address: ]{lizx@mail.nankai.edu.cn}
\affiliation{Nankai University}
\author{Liping Yang}
\email[Electronic address: ]{lpyang@ihep.ac.cn}
\affiliation{Institute of High Energy Physics, Chinese Academy of Sciences}
\author{Bai-Cian Ke}
\email[Electronic address: ]{baiciank@ihep.ac.cn}
\noaffiliation
\author{Liaoyuan Dong}
\email[Electronic address: ]{dongly@ihep.ac.cn}
\affiliation{Institute of High Energy Physics, Chinese Academy of Sciences}
\author{Minggang Zhao}
\email[Electronic address: ]{zhaomg@nankai.edu.cn}
\affiliation{Nankai University}
\begin{abstract}
Finite detector resolution can bias resonance parameters and fit fractions extracted from amplitude analyses. We present the paired Monte Carlo intensity-ratio method for incorporating resolution effects into amplitude analysis. The method approximately accounts for detector resolution by reweighting the Monte Carlo normalization with ratios of reference intensities evaluated at paired truth-level and reconstructed kinematics. A resampling study of simulated $D^+\to K^+\pi^+\pi^-\pi^0$ decays shows that the method substantially reduces resolution-induced shifts in fitted resonance widths. Likelihood evaluation uses intensity calculations and normalization sums, avoiding a response convolution at each data event.
\end{abstract}
\maketitle
\allowdisplaybreaks[1]

\section{Introduction}\label{sec:intro}

Amplitude analysis connects multibody decay distributions to the quantum dynamics of hadrons. Coherent contributions and their interference encode resonance masses and widths, angular structure, and the relative magnitudes and phases of decay amplitudes~\cite{asner2010,zhang2026}. These quantities provide a quantitative basis for hadron spectroscopy and for identifying overlapping contributions in complex final states. Their reliable extraction requires an accurate description of how the underlying decay dynamics are reflected in the measured event distribution.

Finite detector resolution smears reconstructed kinematic distributions and can shift or broaden resonance structures. It reflects the limited precision of tracking and calorimetric measurements, including fluctuations in interactions with detector material and electronic noise. The resulting kinematic migration can involve non-Gaussian tails and correlations between reconstructed variables. Neglecting these effects in amplitude analysis can bias the extracted resonance masses and widths, fit fractions, and relative amplitude magnitudes and phases, especially for narrow resonances.

Large data samples support increasingly detailed studies of multibody decays. As statistical uncertainties decrease, extracting resonance properties and interference parameters requires a likelihood that accurately connects the dynamical intensity to the selected reconstructed distribution. Directly evaluating the multidimensional response integral at each data event and trial parameter point can be computationally demanding. This motivates a formulation that incorporates the response while allowing efficient normalization and likelihood evaluation.

Resolution treatments in amplitude analyses depend on the relative scales of intrinsic widths and detector smearing. Resolution has been neglected when it is small compared with resonance widths~\cite{lhcb2015,lhcb2025}. In searches for narrow states, Gaussian line shapes have instead been used assuming a negligible intrinsic width~\cite{lhcb2023}. Neither limiting approximation generally applies when the two scales are comparable. Explicit treatments include convolution along a selected invariant-mass-squared variable at fixed remaining kinematics~\cite{babar2009} and a two-dimensional response for migration from misreconstructed signal events~\cite{babar2011}. Extending such descriptions to correlated migration in higher-multiplicity decays requires an explicit multidimensional response model.

Monte Carlo (MC) methods offer an alternative to explicit response parameterization by incorporating acceptance and smearing through simulated events~\cite{schmidt1993}. Truth-level MC reweighting has been used to predict reconstructed Dalitz distributions in binned fits~\cite{na482010}, while local MC shifts have been used to construct data-event weights~\cite{xiang2026}. These approaches require binning or local intensity averages around individual data events, respectively. Related methods such as SwdFold use optimal-transport reweighting for multidimensional unfolding~\cite{pan2024}, which introduces a separate optimization of the reweighting function. For amplitude analysis, reconstructed-level intensity ratios have also been used in signal-MC normalization~\cite{zhang2026}. These developments motivate using paired simulated kinematics directly in the MC normalization of an unbinned amplitude likelihood.

We formulate the paired Monte Carlo intensity-ratio (PMIR) method for signal amplitude analysis. We approximate the ratio of the response-convolved trial and reference intensities by the corresponding unconvolved intensity ratio evaluated at reconstructed kinematics. Applying the same approximation to the Monte Carlo normalization gives a signal-shape likelihood that can be updated through successive reference fits. The relevant smoothness condition concerns the trial-to-reference intensity ratio; a shared narrow structure may make this ratio smoother. A resampling study of simulated $D^+\to K^+\pi^+\pi^-\pi^0$ decays examines how two reference updates affect the distributions of fitted resonance parameters and fit fractions.

\section{Exact Signal Density with Detector Resolution}\label{sec:exact}

For a signal sample, the likelihood is constructed from the signal probability density function (PDF), which describes the selected reconstructed-level (rec-level) kinematics. Its construction combines the truth-level decay intensity with the detector response and final event selection. Table~\ref{tab:t1} summarizes the notation used below.

\begin{table*}[t]
\centering
\caption{Notation for the signal density and Monte Carlo normalization.}
\label{tab:t1}
\footnotesize
\begin{tabular}{@{}ll@{}}
\toprule
Symbol & Meaning \\
\midrule
$p^{\mathrm{truth}}$, $p^{\mathrm{rec}}$ & Truth-level and reconstructed (rec-level) final-state kinematics. \\
$d\Phi_n(p^{\mathrm{truth}})$, $dp^{\mathrm{rec}}$ & Truth-level phase-space measure and reconstructed integration measure. \\
$n$, $\Phi_n$ & Final-state multiplicity and phase-space volume, $\Phi_n=\int d\Phi_n(p^{\mathrm{truth}})$. \\
$\theta$ & Model parameters: relative amplitude magnitudes and phases, and resonance masses and widths. \\
$\mathcal M_\theta$, $I_\theta$ & Coherent amplitude and intensity, $I_\theta=\lvert\mathcal M_\theta\rvert^2$. \\
$T(p^{\mathrm{rec}}\mid p^{\mathrm{truth}})$ & Conditional detector-response density, including measurement and reconstruction. \\
$C(p^{\mathrm{rec}})$ & Final-selection indicator (including candidate selection): 1 if accepted, 0 otherwise. \\
$\epsilon(p^{\mathrm{truth}})$ & Probability that a truth-level event enters the selected sample. \\
$G_\theta(p^{\mathrm{rec}})$ & Reconstructed signal intensity after detector response. \\
$f_S(p^{\mathrm{rec}};\theta)$, $N_S(\theta)$ & Selected signal PDF and its normalization. \\
$N_D$, $N_P$ & Selected signal-data count and total generated MC count. \\
$i$, $k$ & Data-event and simulated-event indices. \\
\bottomrule
\end{tabular}
\end{table*}

\begin{figure*}[t]
\centering
\includegraphics[width=0.84\textwidth]{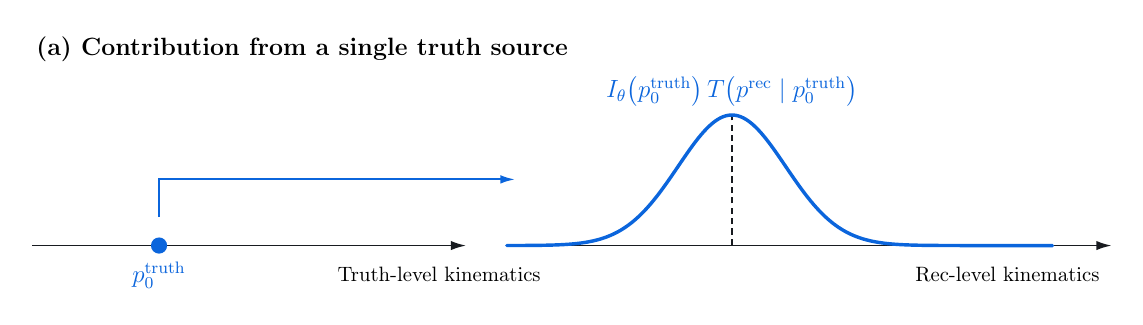}\par\smallskip
\includegraphics[width=0.84\textwidth]{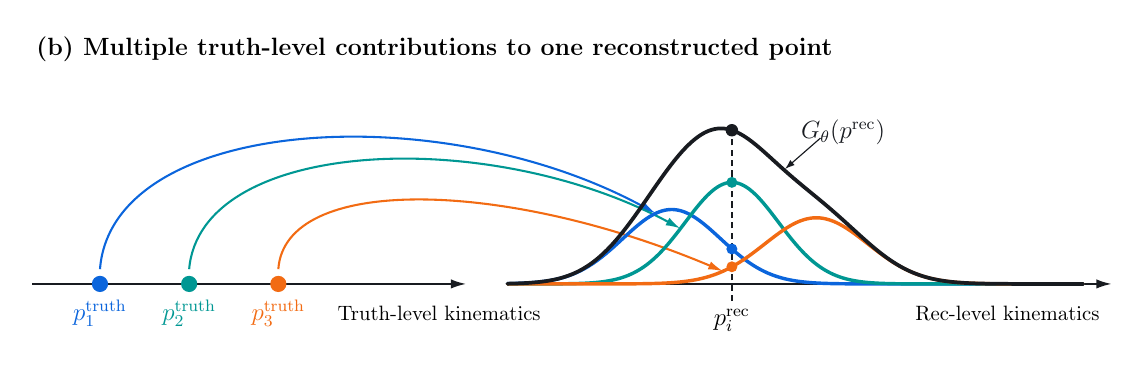}
\caption{Detector migration in the signal density. (a) The contribution from fixed truth-level kinematics. (b) Contributions from multiple truth-level points to a given reconstructed point.}
\label{fig:f1}
\end{figure*}

At fixed truth-level kinematics, detector effects produce a distribution at rec level [Fig.~\ref{fig:f1}(a)]. Conversely, a given rec-level point receives contributions from a range of truth-level kinematics [Fig.~\ref{fig:f1}(b)]. Integrating $I_\theta T$ over truth-level phase space gives the rec-level signal intensity in Eq.~\eqref{eq:response-intensity}. Selection and normalization then give the signal PDF in Eq.~\eqref{eq:exact-signal-pdf}:

{\small
\begin{align}
G_\theta(p^{\mathrm{rec}})
&=\int d\Phi_n(p^{\mathrm{truth}})\,I_\theta(p^{\mathrm{truth}})
  T(p^{\mathrm{rec}}\mid p^{\mathrm{truth}}),
\label{eq:response-intensity}\\[6pt]
f_S(p^{\mathrm{rec}};\theta)
&=\frac{C(p^{\mathrm{rec}})G_\theta(p^{\mathrm{rec}})}
{\displaystyle\int dp^{\mathrm{rec}\prime}\,C(p^{\mathrm{rec}\prime})G_\theta(p^{\mathrm{rec}\prime})}.
\label{eq:exact-signal-pdf}
\end{align}
}

The normalization $N_S(\theta)$ in the denominator of Eq.~\eqref{eq:exact-signal-pdf} can be evaluated by Monte Carlo integration. At fixed truth-level kinematics, integrating $T$ over accepted rec-level outcomes gives the selection probability $\epsilon(p^{\mathrm{truth}})$. Uniformly generated phase-space events passed through detector simulation, reconstruction, and selection incorporate this efficiency through the accepted sample. For selected data events, $C(p^{\mathrm{rec}})=1$. Substituting $G_\theta$ into the denominator and interchanging the integrations gives an integral over truth-level phase space. Estimating this integral with the selected phase-space MC sample gives Eq.~\eqref{eq:exact-mc-pdf}:

\begin{widetext}
{\small
\begin{equation}
\label{eq:exact-mc-pdf}
\begin{aligned}
f_S(p^{\mathrm{rec}};\theta)
&=\frac{G_\theta(p^{\mathrm{rec}})}
{\displaystyle\int dp^{\mathrm{rec}\prime}\,C(p^{\mathrm{rec}\prime})
  \int d\Phi_n(p^{\mathrm{truth}})\,I_\theta(p^{\mathrm{truth}})
  T(p^{\mathrm{rec}\prime}\mid p^{\mathrm{truth}})}\\[6pt]
&=\frac{G_\theta(p^{\mathrm{rec}})}
{\displaystyle\int d\Phi_n(p^{\mathrm{truth}})\,I_\theta(p^{\mathrm{truth}})
  \underbrace{\int dp^{\mathrm{rec}\prime}\,C(p^{\mathrm{rec}\prime})
  T(p^{\mathrm{rec}\prime}\mid p^{\mathrm{truth}})}_{\epsilon(p^{\mathrm{truth}})}}\\[6pt]
&\simeq\frac{G_\theta(p^{\mathrm{rec}})}
{\displaystyle\frac{\Phi_n}{N_P}
  \sum_{k\in\mathrm{selected}}I_\theta(p_k^{\mathrm{truth}})}.
\end{aligned}
\end{equation}
}
\end{widetext}

When detector smearing is much smaller than the intrinsic resonance widths, resolution effects can be neglected by setting $p^{\mathrm{rec}}=p^{\mathrm{truth}}$. Within this approximation, the conventional phase-space PDF widely used in amplitude analyses and its finite-MC estimate are

{\small
\begin{equation}
\label{eq:zero-resolution}
\begin{aligned}
f_0(p^{\mathrm{rec}};\theta)
&=\frac{\epsilon(p^{\mathrm{rec}})I_\theta(p^{\mathrm{rec}})}
{\displaystyle\int d\Phi_n(p^{\mathrm{rec}\prime})\,\epsilon(p^{\mathrm{rec}\prime})I_\theta(p^{\mathrm{rec}\prime})}\\[6pt]
&\simeq\frac{\epsilon(p^{\mathrm{rec}})I_\theta(p^{\mathrm{rec}})}
{\displaystyle\frac{\Phi_n}{N_P}\sum_{k\in\mathrm{selected}}I_\theta(p_k^{\mathrm{rec}})}.
\end{aligned}
\end{equation}
}

\section{Likelihood with the Local Intensity-Ratio Approximation}\label{sec:approximation}

The exact PDF in Sec.~\ref{sec:exact} requires an accurate multidimensional detector response $T$ and a response integral at each data event whenever the trial parameters change. Both requirements are difficult to meet in multibody amplitude analyses. The PMIR method starts from a fixed reference amplitude model that approximately describes the data and tracks changes during the fit through trial-to-reference intensity ratios. Paired truth-level and rec-level simulated kinematics carry the migration information into the event weights, allowing resolution effects to enter the fit without evaluating $T$ explicitly.

The reference parameters $\theta_0$ are chosen in advance and held fixed during each fit. They can initially be taken from a fit that neglects resolution, as described in Sec.~\ref{sec:exact}. The trial model has parameters $\theta$ that are varied to find the best description of the data, and its intensity is $I_\theta$. For the fixed reference model, the truth-level intensity $I_0$ and rec-level signal intensity $G_0$ are defined in Eqs.~\eqref{eq:reference-intensity} and \eqref{eq:reference-response}:

{\small
\begin{align}
I_0(p^{\mathrm{truth}})
&\equiv I_{\theta_0}(p^{\mathrm{truth}}),
\label{eq:reference-intensity}\\[6pt]
G_0(p^{\mathrm{rec}})
&=\int d\Phi_n(p^{\mathrm{truth}})\,I_0(p^{\mathrm{truth}})
  T(p^{\mathrm{rec}}\mid p^{\mathrm{truth}}).
\label{eq:reference-response}
\end{align}
}

For $G_0(p^{\mathrm{rec}})>0$, we substitute $I_\theta=I_0(I_\theta/I_0)$ into Eq.~\eqref{eq:response-intensity}. If the intensity ratio varies slowly over the contributing truth-level kinematics, we approximate it by its value at $p^{\mathrm{rec}}$. Taking this value outside the integral gives the approximation in Eq.~\eqref{eq:intensity-ratio}:

\begin{widetext}
{\small
\begin{equation}
\label{eq:intensity-ratio}
\begin{aligned}
\frac{G_\theta(p^{\mathrm{rec}})}{G_0(p^{\mathrm{rec}})}
&=\frac{\displaystyle\int d\Phi_n(p^{\mathrm{truth}})\,I_0(p^{\mathrm{truth}})
  T(p^{\mathrm{rec}}\mid p^{\mathrm{truth}})
  \frac{I_\theta(p^{\mathrm{truth}})}{I_0(p^{\mathrm{truth}})}}
{\displaystyle\int d\Phi_n(p^{\mathrm{truth}})\,I_0(p^{\mathrm{truth}})
  T(p^{\mathrm{rec}}\mid p^{\mathrm{truth}})}\\[6pt]
&\approx\frac{\displaystyle\frac{I_\theta(p^{\mathrm{rec}})}{I_0(p^{\mathrm{rec}})}
  \int d\Phi_n(p^{\mathrm{truth}})\,I_0(p^{\mathrm{truth}})
  T(p^{\mathrm{rec}}\mid p^{\mathrm{truth}})}
{\displaystyle\int d\Phi_n(p^{\mathrm{truth}})\,I_0(p^{\mathrm{truth}})
  T(p^{\mathrm{rec}}\mid p^{\mathrm{truth}})}\\[6pt]
&=\frac{I_\theta(p^{\mathrm{rec}})}{I_0(p^{\mathrm{rec}})}.
\end{aligned}
\end{equation}
}
\end{widetext}

The corresponding approximate rec-level signal intensity is
\begingroup
\setlength{\abovedisplayskip}{6pt}
\setlength{\belowdisplayskip}{6pt}
\setlength{\abovedisplayshortskip}{4pt}
\setlength{\belowdisplayshortskip}{6pt}
\begin{equation}
\label{eq:approximate-intensity}
\widetilde G_\theta(p^{\mathrm{rec}})
=G_0(p^{\mathrm{rec}})\frac{I_\theta(p^{\mathrm{rec}})}{I_0(p^{\mathrm{rec}})}.
\end{equation}
\endgroup

Common resonance structures can make the intensity ratio smoother than either intensity. Equation~\eqref{eq:approximate-intensity} reproduces the reference signal intensity at $\theta=\theta_0$ and is exact in the zero-migration limit. Normalizing this intensity over the selected region gives the approximate signal PDF in Eq.~\eqref{eq:approximate-pdf}:

\begin{widetext}
{\small
\begin{equation}
\label{eq:approximate-pdf}
\begin{gathered}
\widetilde f_S(p^{\mathrm{rec}};\theta)
=\frac{C(p^{\mathrm{rec}})\widetilde G_\theta(p^{\mathrm{rec}})}
{\displaystyle\int dp^{\mathrm{rec}\prime}\,C(p^{\mathrm{rec}\prime})\widetilde G_\theta(p^{\mathrm{rec}\prime})}\\[6pt]
{}=\frac{C(p^{\mathrm{rec}})G_0(p^{\mathrm{rec}})
  \dfrac{I_\theta(p^{\mathrm{rec}})}{I_0(p^{\mathrm{rec}})}}
{\displaystyle\int dp^{\mathrm{rec}\prime}\,C(p^{\mathrm{rec}\prime})G_0(p^{\mathrm{rec}\prime})
  \frac{I_\theta(p^{\mathrm{rec}\prime})}{I_0(p^{\mathrm{rec}\prime})}}.
\end{gathered}
\end{equation}
}
\end{widetext}

For a fixed reference, $C(p_i^{\mathrm{rec}})G_0(p_i^{\mathrm{rec}})/I_0(p_i^{\mathrm{rec}})$ is independent of the trial parameters. In the signal log likelihood it contributes an additive constant, so $G_0$ need not be evaluated at each data event. The denominator, $\widetilde N_S(\theta)$, can be evaluated with paired simulation as follows.

Substituting the response integral for $G_0$ into the denominator and interchanging the integrations yields an outer integral over truth-level phase space and an inner integral over rec-level outcomes. In uniform phase-space simulation, the generated truth-level kinematics sample the outer integral, while detector simulation, reconstruction, and selection sample the inner integral over rec-level outcomes. The resulting MC sum in Eq.~\eqref{eq:paired-normalization} uses the reference intensity at truth level and the trial-to-reference intensity ratio at rec level:

\begin{widetext}
{\small
\begin{equation}
\label{eq:paired-normalization}
\begin{aligned}
\widetilde N_S(\theta)
&=\int dp^{\mathrm{rec}}\,C(p^{\mathrm{rec}})G_0(p^{\mathrm{rec}})
  \frac{I_\theta(p^{\mathrm{rec}})}{I_0(p^{\mathrm{rec}})}\\[6pt]
&=\int d\Phi_n(p^{\mathrm{truth}})\,I_0(p^{\mathrm{truth}})
  \int dp^{\mathrm{rec}}\,C(p^{\mathrm{rec}})
  T(p^{\mathrm{rec}}\mid p^{\mathrm{truth}})
  \frac{I_\theta(p^{\mathrm{rec}})}{I_0(p^{\mathrm{rec}})}\\[6pt]
&\simeq\frac{\Phi_n}{N_P}\sum_{k\in\mathrm{selected}}
  \frac{I_0(p_k^{\mathrm{truth}})}{I_0(p_k^{\mathrm{rec}})}
  I_\theta(p_k^{\mathrm{rec}}).
\end{aligned}
\end{equation}
}

With the reference model held fixed, substituting this MC estimate into the signal log likelihood gives

{\small
\begin{equation}
\label{eq:paired-likelihood}
\widehat{\widetilde\ell}_S(\theta)
=\sum_{i=1}^{N_D}\log I_\theta(p_i^{\mathrm{rec}})
-N_D\log\left[
  \frac{\Phi_n}{N_P}\sum_{k\in\mathrm{selected}}
  \frac{I_0(p_k^{\mathrm{truth}})}{I_0(p_k^{\mathrm{rec}})}
  I_\theta(p_k^{\mathrm{rec}})
\right].
\end{equation}
}
\end{widetext}

\begin{table*}[t]
\centering
\caption{Fitted resonance masses and widths, given as means $\pm$ standard deviations over 500 resampled data sets.}
\label{tab:t2}
\footnotesize
\setlength{\tabcolsep}{2.5pt}
\begin{tabular}{@{}lrrrr@{}}
\toprule
Quantity & Input & R0 & R1 & R2 \\
\midrule
$m_{K^{*0}}$ (MeV/$c^2$) & 895.55 & $895.67\pm0.87$ & $895.53\pm0.77$ & $895.52\pm0.77$ \\
$\Gamma_{K^{*0}}$ (MeV) & 47.30 & $49.10\pm2.15$ & $47.22\pm1.85$ & $47.20\pm1.83$ \\
$m_{\rho^+}$ (MeV/$c^2$) & 775.11 & $774.85\pm2.03$ & $775.10\pm1.85$ & $775.11\pm1.84$ \\
$\Gamma_{\rho^+}$ (MeV) & 149.10 & $151.56\pm5.49$ & $149.46\pm4.80$ & $149.46\pm4.78$ \\
$m_\omega$ (MeV/$c^2$) & 782.66 & $782.11\pm0.39$ & $782.63\pm0.23$ & $782.63\pm0.23$ \\
$\Gamma_\omega$ (MeV) & 8.68 & $12.27\pm0.55$ & $8.88\pm0.44$ & $8.88\pm0.44$ \\
$m_\phi$ (MeV/$c^2$) & 1019.46 & $1019.15\pm0.30$ & $1019.46\pm0.11$ & $1019.46\pm0.11$ \\
$\Gamma_\phi$ (MeV) & 4.25 & $8.11\pm0.40$ & $4.40\pm0.24$ & $4.41\pm0.24$ \\
\bottomrule
\end{tabular}
\end{table*}

\begin{table}[t]
\centering
\caption{Truth-level fit fractions in percent, with the same convention as Table~\ref{tab:t2}.}
\label{tab:t3}
\footnotesize
\setlength{\tabcolsep}{2.5pt}
\begin{tabular}{@{}lrrrr@{}}
\toprule
Quantity & Input & R0 & R1 & R2 \\
\midrule
$F(K^{*0}\rho^+)$ & 49.86 & $50.34\pm0.96$ & $50.13\pm0.97$ & $50.11\pm0.97$ \\
$F(K^+\omega)$ & 25.53 & $25.62\pm0.80$ & $25.58\pm0.80$ & $25.57\pm0.80$ \\
$F(K^+\phi)$ & 24.08 & $23.54\pm0.78$ & $23.78\pm0.78$ & $23.81\pm0.78$ \\
\bottomrule
\end{tabular}
\end{table}

Together, the local intensity-ratio approximation and paired simulated events yield a signal likelihood that accounts for detector resolution without evaluating the response integral at each data event. The fitted parameters can then be used to update the reference model, recalculate the reference intensities, and repeat the fit. Bringing the reference closer to the fitted model can make the intensity ratio vary more slowly over the contributing truth-level region and improve the local approximation.

The paired phase-space MC sum in Eq.~\eqref{eq:paired-normalization} estimates the same normalization integral as the signal-MC integration used in amplitude analyses~\cite{besiii2021kkpi,besiii2026ks00}. After detector simulation, reconstruction, and selection, signal MC generated from the fixed reference intensity $I_0(p^{\mathrm{truth}})$ has a rec-level density proportional to $C(p^{\mathrm{rec}})G_0(p^{\mathrm{rec}})$. This sample gives the normalization estimate in Eq.~\eqref{eq:signal-mc-normalization}:

{\small
\begin{equation}
\label{eq:signal-mc-normalization}
\begin{aligned}
\widetilde N_S(\theta)
&=\int dp^{\mathrm{rec}}\,C(p^{\mathrm{rec}})G_0(p^{\mathrm{rec}})
  \frac{I_\theta(p^{\mathrm{rec}})}{I_0(p^{\mathrm{rec}})}\\[6pt]
&\simeq\frac{N_S(\theta_0)}{N_{0,\mathrm{sel}}}
  \sum_{k=1}^{N_{0,\mathrm{sel}}}
  \frac{I_\theta(p_k^{\mathrm{rec}})}{I_0(p_k^{\mathrm{rec}})},
\end{aligned}
\end{equation}
}

where $N_{0,\mathrm{sel}}$ is the selected signal-MC count and $N_S(\theta_0)$ is independent of the trial parameters. The narrow-resonance width check reported in Ref.~\cite{besiii2025fivepi} supports the effectiveness of this resolution treatment.

\section{Validation with Simulated Experiments}\label{sec:validation}

To test PMIR, we generated a large signal MC sample of $D^+\to K^+\pi^+\pi^-\pi^0$ decays containing coherent $K^{*0}\rho^+$, $K^+\omega$, and $K^+\phi$ components with target relative fractions of approximately $2:1:1$. The generated events were processed through simulation, reconstruction, and event selection. Five hundred data sets, each containing $3{,}000$ rec-level events, were drawn with replacement from the resulting selected signal sample. The normalization used uniform phase-space MC events with paired truth-level and rec-level kinematics.

The comparison began with a fit using the zero-resolution PDF in Eq.~\eqref{eq:zero-resolution}, evaluating both data and MC intensities at rec level (R0). Its fitted amplitude coefficients then defined the reference model for a second fit (R1), performed by maximizing Eq.~\eqref{eq:paired-likelihood}. Updating the reference with the R1 coefficients and repeating the fit gave R2. The reference masses and widths remained fixed at their input values throughout.

The main fits varied only the amplitude coefficients. Separate diagnostic fits were used to assess the effect of PMIR on the extracted resonance masses and widths. Each diagnostic fit varied one mass or width together with the amplitude coefficients, while the remaining masses and widths stayed fixed. The results were compared with the input values.

\begin{figure}[t]
\centering
\includegraphics[width=\columnwidth,pagebox=cropbox]{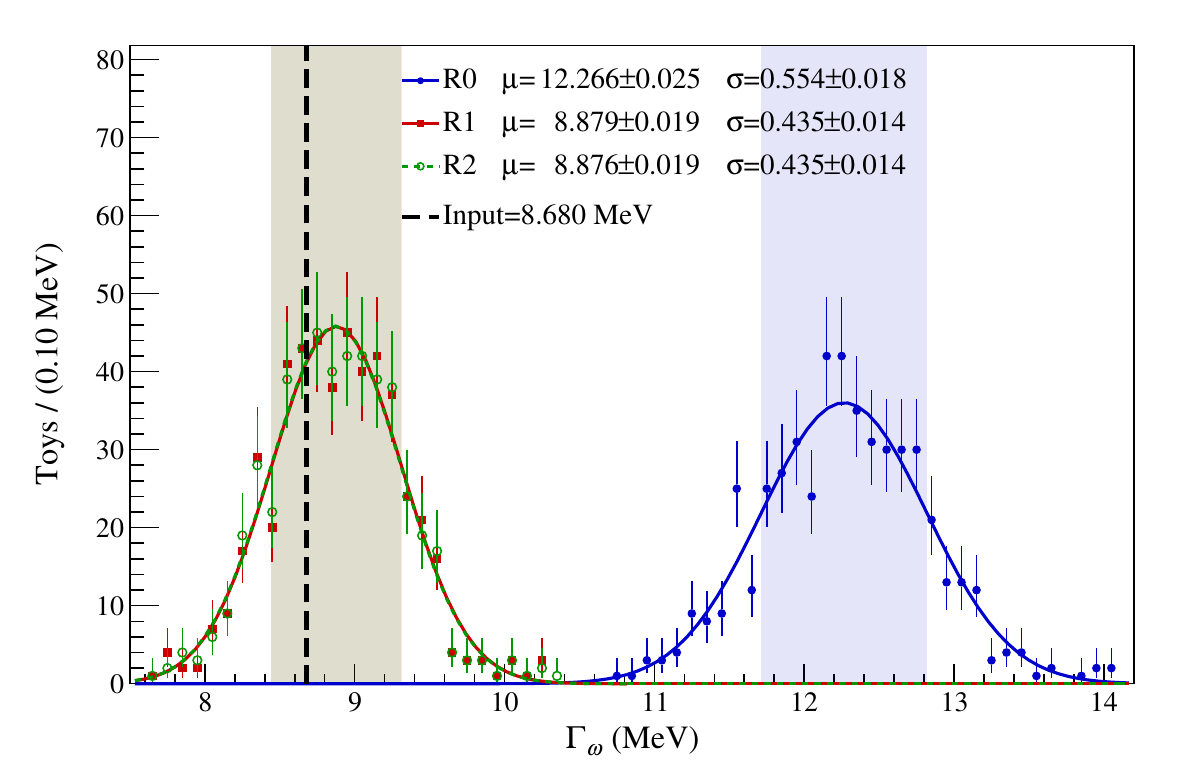}\par
\makebox[\columnwidth][c]{(a)}\par\smallskip
\includegraphics[width=\columnwidth,pagebox=cropbox]{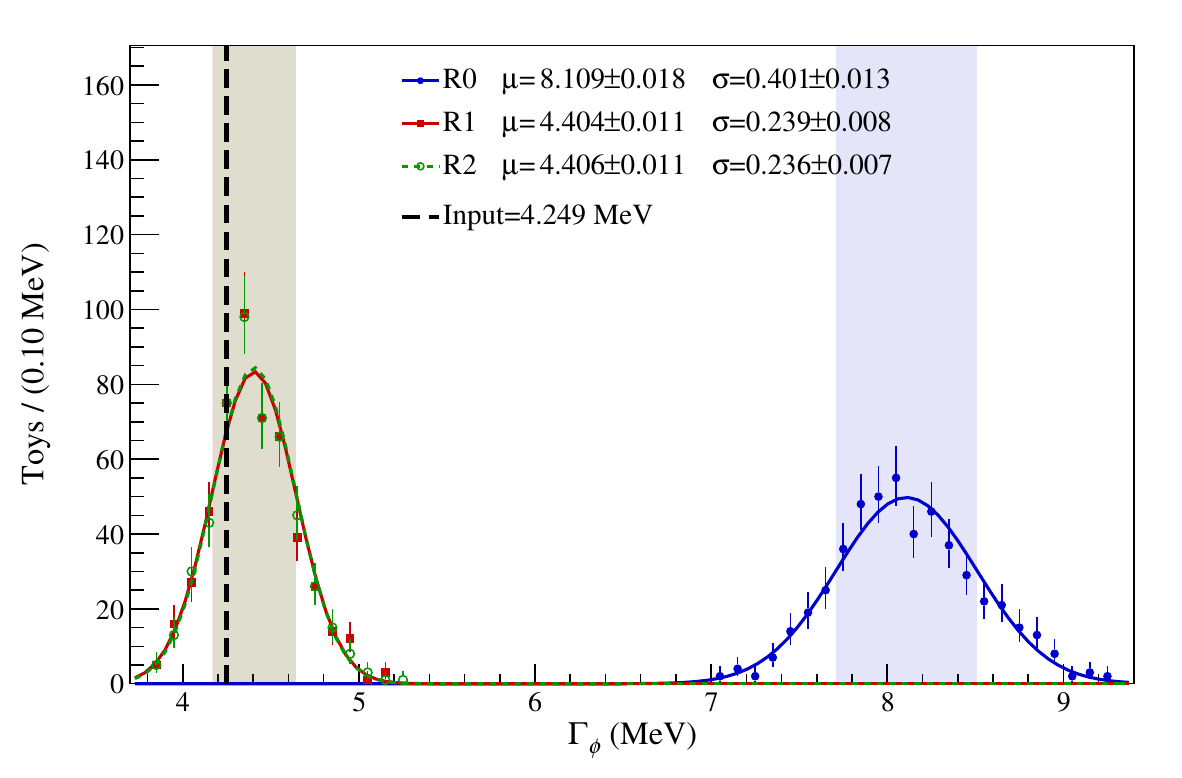}\par
\makebox[\columnwidth][c]{(b)}\par
\caption{Distributions of the fitted (a) $\omega$ and (b) $\phi$ widths before and after reference updates. Curves represent Gaussian fits, shaded bands span one standard deviation about the ensemble means, and black dashed lines mark the inputs.}
\label{fig:f2}
\end{figure}

The fitted masses and widths in Table~\ref{tab:t2} are reported as means $\pm$ standard deviations across the 500 resampled data sets. At R0, the mean fitted $\omega$ and $\phi$ widths substantially exceeded the input values. With the reference masses and widths fixed to their generated values, the R1 and R2 means were within one ensemble standard deviation of the inputs. The second update produced only small changes (Fig.~\ref{fig:f2}). %The legend uncertainties are the statistical errors on the ensemble means and standard deviations.

\begin{figure*}[t]
\centering
\begin{minipage}[t]{0.49\textwidth}
\centering
\includegraphics[width=\linewidth,pagebox=cropbox]{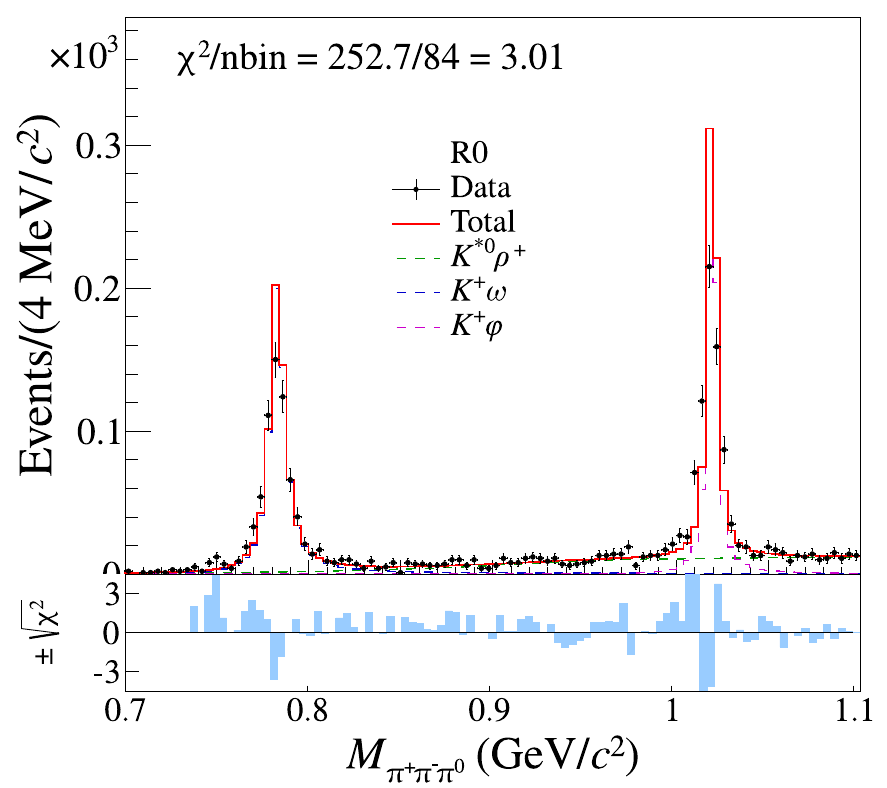}\par
\makebox[\linewidth][c]{(a)}
\end{minipage}\hfill
\begin{minipage}[t]{0.49\textwidth}
\centering
\includegraphics[width=\linewidth,pagebox=cropbox]{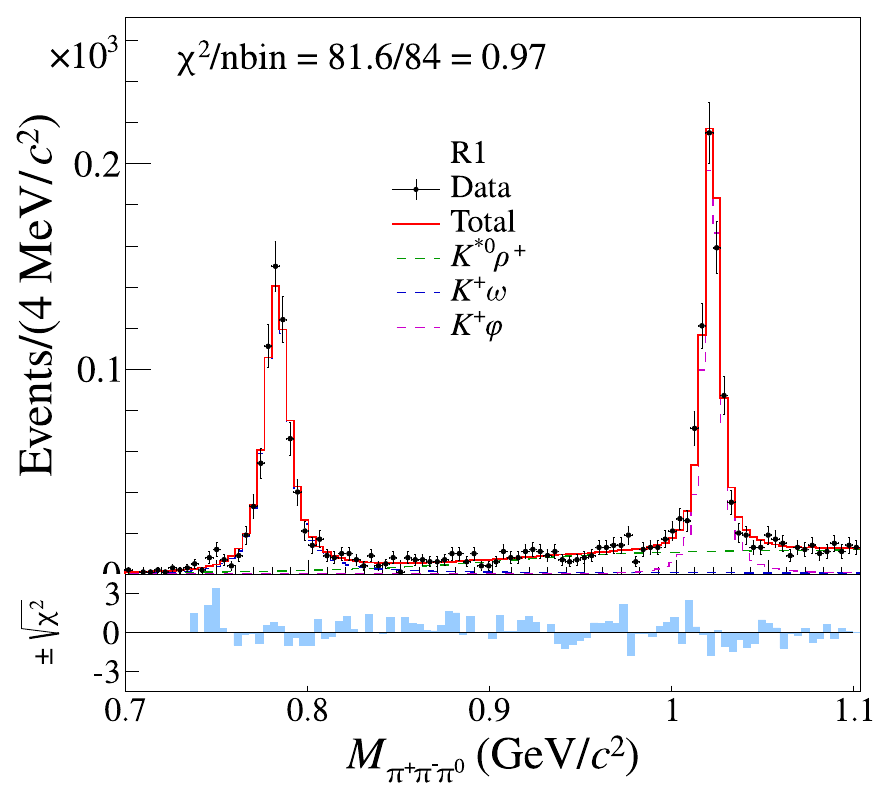}\par
\makebox[\linewidth][c]{(b)}
\end{minipage}
\caption{Reconstructed $M(\pi^+\pi^-\pi^0)$ projections for the same resampled data set using (a) R0 and (b) R1. Black points show the simulated data, red histograms the total fitted distributions, and dashed histograms the individual components. Lower panels show the signed square roots of the binwise $\chi^2$ contributions.}
\label{fig:f3}
\end{figure*}

Figure~\ref{fig:f3} compares the reconstructed $M(\pi^+\pi^-\pi^0)$ spectra for the same resampled data set at R0 and R1. The PMIR method substantially improves the description of the mass spectrum around the $\omega$ and $\phi$ peaks.

Table~\ref{tab:t3} summarizes the truth-level fit fractions. For each fitted model, all component fit fractions are normalized to the total intensity integrated over truth-level phase space. For all three components, the mean fit fractions at R0, R1, and R2 are within one ensemble standard deviation of the inputs. The reference updates brought the $K^{*0}\rho^+$ and $K^+\phi$ means closer to their input values.

\section{Summary and Conclusions}\label{sec:conclusion}

We have presented the PMIR method for incorporating detector resolution into signal amplitude analysis through a local intensity-ratio approximation. Paired truth-level and reconstructed MC kinematics carry multidimensional migration and correlation information into the likelihood normalization. For a fixed reference model, the reference-dependent MC weights can be precomputed, so each likelihood evaluation requires intensity calculations and a weighted normalization sum. The paired phase-space MC sample can be reused across reference updates. This construction retains an unbinned likelihood without an explicit analytic response density or a response convolution at each data event.

In the simulated $D^+\to K^+\pi^+\pi^-\pi^0$ study, PMIR substantially reduced the $\omega$ and $\phi$ width shifts in the separate diagnostic fits and improved the reconstructed three-pion mass spectrum, while retaining fit fractions consistent with the inputs.

The formulation uses model intensities and paired simulated kinematics, so it can be incorporated into existing unbinned fitting frameworks through modified MC normalization weights. It accommodates different final-state multiplicities and amplitude parameterizations when the detector simulation describes the relevant migration and the trial-to-reference intensity ratio varies slowly over the contributing truth-level region. Under these conditions, PMIR can incorporate resolution effects in multibody amplitude analyses involving narrow resonances or correlated detector smearing.

\begin{acknowledgments}
This work was supported by the National Natural Science Foundation of China (NSFC) under Grant Nos.~12192262, 12635006, 12675104 and 12192263, the Excellent Youth Foundation of Henan Scientific Committee under Contract No.~242300421044, and the Joint Large-Scale Scientific Facility Fund of the NSFC and the Chinese Academy of Sciences under Contract No.~U2032104.
\end{acknowledgments}

\bibliography{references}
\end{document}